\documentclass[12pt]{article}
\usepackage{graphicx}
\usepackage{color}

\definecolor{darkgreen}{rgb}{0.0,0.35,0.0}
\begin{document}
\title{Molecular Dynamics Simulations of the Thermal 
Glass Transition in Polymer Melts: $\alpha$-Relaxation Behaviour}

\author{Christoph Bennemann\\
	Wolfgang Paul\\
	Kurt Binder\\
Institut f\"ur Physik\\
        55099 Mainz\\
	Germany
\and
	Burkhard D\"unweg\\
MPI f\"ur Polymerforschung \\
	Ackermannweg 10 \\
	55128 Mainz\\
	Germany}
\maketitle
\begin{abstract}
We present Molecular Dynamics simulations of the thermal glass transition
in a dense model polymer liquid. We performed a comparative study of both
constant volume and constant pressure cooling of the polymer melt. Great
emphasis was laid on a careful equilibration of the dense polymer melt 
at all studied temperatures. Our model introduces competing length 
scales in the interaction to prevent any crystallisation tendency. In
this first manuscript we analyse the structural properties as a function
of temperature and the long time or $\alpha$-relaxation behaviour as
observed in the dynamic structure factor and the self-diffusion of the 
polymer chains. The $\alpha$-relaxation can be consistently analysed in
terms of the mode coupling theory (MCT) of the glass transition. The
mode coupling critical temperature, $T_c$, and the exponent, $\gamma$,
defining the power law divergence of the $\alpha$-relaxation timescale
both depend on the thermodynamic ensemble employed in the simulation.
\end{abstract}

\section{Introduction}
Understanding the glass transition in supercooled materials is at
the same time a great challenge in condensed matter theory \cite{jaeckle,goetze1} and of
high technological importance \cite{technik}. Polymers constitute a class of materials
with a very small crystallisation tendency. The ubiquity of amorphous
polymeric materials has made them a longstanding focus of the
experimental characterisation of the glass transition \cite{McKenna} as well
as efforts to derive models of this transition \cite{DiMarzio}. 
Traditionally this work focused on the 
temperature range in which the typical relaxation times in the material
are macroscopic, i.e. in the range of seconds.

With the development of the mode coupling theory of the glass transition
 \cite{goetze1,sjlander,goetze2,goetze3,leutheusser}
the interest was shifted to the temperature regime of the undercooled
liquid and relaxation times in the ns to $\mu$s range. This ignited
a tremendous effort over the last decade \cite{Heraklion,Alicante,Vigo}
on the side of experiment and
computer simulations to test the predictions and range of validity
of this theory on all kinds of glass forming materials. Originally
this theory has been developed for hard sphere liquids but it has
been applied to and claimed to have been tested on as diverse materials as
colloids \cite{puce,vanMegen}, 
ionic glasses \cite{mezei}, 
molecular \cite{roessler} liquids and polymers \cite{Frick}. The emerging
picture seems to be, that the theory can be applied in situations
where the glass transition is determined by the repulsive part
of the intermolecular interactions, i.e. where there are no site specific
attractive interactions which could give rise for instance to
network formation like in $SiO_2$.

On the computer simulation side there has been a very detailed test
of the MCT on a mixture of Lennard-Jones particles \cite{kob2,kob3,Nauroth}
which belongs to
the class of materials discussed above. The glass transition in
polymer melts has been studied in great detail with Monte Carlo
simulations of the bond fluctuation model \cite{paul2,baschnag1,baschnag2,Wolf1,Wolf2,Wolf3,okun} 
which can also be
applied to the modelling of real polymers \cite{paul1}. The combination of
lattice model and Monte Carlo method necessarily means that one in
general studies the glass transition at constant volume and that one has
completely neglected inertia effects in the short time dynamics.
Any motions on scales smaller than a lattice spacing are completely eliminated, of course.
Both of these drawbacks can be remedied by resorting to a Molecular Dynamics
simulation of a continuum model. Work in this direction mainly
used atomistic polyethylene-like models. Early work \cite{brown,roe1} focused
on the glass transition as a phenomenon of macroscopic time
scales, observing the break in the dependence of the specific
volume on temperature. This study as well as later work \cite{troe,roe2} used
high quenching rates loosing one of the main advantages of such
polymeric glass formers, namely the ability to equilibrate
chain conformations and local packing in the amorphous state
without intervening crystallisation tendency. In the series
of works \cite{roe1,troe,roe2} no systematic study of quenching rate effects or the
degree of equilibration at different temperatures was reported
on. Especially for the dynamic structure factor it has been
shown \cite{baschnag1,baschnag2}, however, 
that the observed behaviour depends strongly
on the degree of equilibration one has achieved.

Thus we decided to perform a systematic study of the glass transition
in a polymer melt using a simple coarse-grained polymer model in
the continuum consisting of Lennard-Jones particles connected
by nonlinear springs \cite{kremer}. This model can be equilibrated with respect
to local packing and chain conformations down to temperatures
well in the regime of the undercooled liquid. All our results on the
dynamic properties of the polymer melt are therefore equilibrium
dynamics. We also decided to simulate this model under constant volume (NVT) 
as well as constant pressure (NpT) conditions to study the difference 
and similarities of the glass transition in these ensembles. Since both
cooling methods follow different paths in the state space of the model
we expect quantitative differences between the observations at constant
volume and constant pressure, albeit qualitatively similar behavior,
as was seen experimentally \cite{Koppelmann}. For a direct quantitative
comparison one would need, for instance, a whole set of constant pressure
cooling curves.

The remainder of this paper is organised as follows. In Section \ref{secII}
we will discuss our model and the simulation procedure for
the NVT as well as NpT simulations. Section \ref{secIII} will present a
comparison of the static properties of the melts in the NVT
and NpT simulations. In Section \ref{secIV} we will discuss our results for
the $\alpha$-relaxation dynamics as observed in the dynamic
structure factor and the self-diffusion of the polymer chains
and Section \ref{secV} will present our conclusions.

A detailed analysis of the $\beta$-relaxation regime predicted by
mode coupling theory along the lines of Refs. \cite{baschnag1,baschnag2}
will be presented in
a forthcoming work.

\section{Model and Simulation Technique}\label{secII}

\noindent
For modelling the inter- and intramolecular forces we used
a bead-spring model derived from the one suggested by Kremer 
and Grest \cite{kremer} and also used in several recent simulations
 \cite{duenweg1,duenweg2}. We however included here also the attractive
part of the Lennard-Jones potential, since previous work \cite{Wolf1,Wolf2}
had shown that without such an attraction the model would produce a negative
thermal expansion coefficient.

\noindent
Each chain consisted of 10 beads with mass m set to unity. Between
all monomers there acted a truncated Lennard-Jones potential:

\begin{equation}
U_{LJ}(r_{ij}) = \left\{ \begin{array}{r@{\quad:\quad}l}
4\epsilon \left[ \left(\frac{\sigma}{r_{ij}}\right)^{12}
- \left(\frac{\sigma}{r_{ij}}\right)^{6}\right] + C & r_{ij} < 2 \cdot 2^{\frac{1}{6}} \sigma\\
0 & r_{ij} \geq 2 \cdot 2^{\frac{1}{6}}\sigma 
\end{array} \right. ,
\end{equation}

\noindent
where $C$ was a constant which guaranteed that the potential was continuous everywhere. Since
it was not our aim to simulate a specific polymer we used Lennard-Jones
units where $\epsilon$ and $\sigma$ are set to
unity. Note that this means that all quantities are dimensionless. 
In addition to the Lennard-Jones potential a FENE backbone potential was applied along
the chain:

\begin{equation}
U_F(r_{ij})=-\frac{k}{2}R_{0}^2 \ln(1-(\frac{r_{ij}}{R_0})^2).
\end{equation}

\noindent
The parameters of the potential were set to $k=30$ and $R_0=1.5$ guaranteeing a certain
stiffness of the bonds while avoiding high frequency modes (which would require a rather small
time step for the integration) and chain crossing. Furthermore with these parameters we set the favoured bondlength
to a value slightly lower than the length favoured by the Lennard-Jones potential. Thus
we introduced two different incompatible length scales in our system, which should help to prevent 
the emergence of long range order at lower temperatures.

\noindent
All simulations in the NVT ensemble were performed using a Nos\'e-Hoover thermostat
 \cite{Nose1,Hoover1} to keep the temperature at the desired level. In this technique the
 model system
is coupled to a heat bath which represents an additional degree of freedom represented
by the variable $\zeta$. The equations of motion are

\begin{eqnarray}
\frac{d{\bf q}_i}{dt} & = &\frac{{\bf p}_i}{m_i}\\
\frac{d{\bf p}_i}{dt} & = &{\bf F}_i - \zeta{\bf p}_i\\
\frac{d\zeta}{dt} & = & \frac{1}{Q}(\sum_i \frac{{\bf p}_i^2}{m_i} - g\:k_B\: T)\:,
\end{eqnarray}

\noindent
where ${\bf F}_i$ is the total force acting on particle $i$ due to the potentials described above
and $Q$ represents the mass of the heat bath, while $g$ is the number of degrees of freedom. 
Note that $\zeta$ fluctuates around zero and can
thus become negative. The mass $Q$ has to be chosen with great care 
 \cite{Hoover2,Tolla} since otherwise one may not obtain a canonical distribution. If for example
$Q$ is very large, the kinetic energy and therefore the temperature starts to oscillate with
an undesired large amplitude. Instead of a canonical distribution one then obtains a two-peaked 
distribution. In principle any problems could be avoided by using a chain of thermostats 
 \cite{Klein1} but that would have worsened the computational effort and was thus discarded.
For optimal results the intrinsic frequency of the heat bath should be approximately
equal to the intrinsic frequency of the model system of which a theoretical estimate
was obtained by calculating the frequency of a particle in a fcc lattice subjected to 
Lennard-Jones potentials. The intrinsic frequency of the heat bath is given by \cite{Tolla}:

\begin{equation}
\frac{1}{\omega_s} = 2\pi\sqrt{\frac{Q}{2 g k_B T}}.
\end{equation}

Setting $\omega_s$ 
equal to the theoretically obtained frequency and rearranging this equation yields 
an expression for $Q$. Note that $Q$ depends explicitly on the temperature and
therefore had to be adjusted for every simulation temperature. 
During all simulations no suspicious behaviour due to the 
choice of $Q$ was observed. We also performed several Monte Carlo
simulations using both a continuum configurational bias method (CCB) \cite{Siepman,pablo1} 
and so called smart reptation which were carried out
at the temperature $T=1.0$ in order to check the validity of the Molecular Dynamics
algorithm and to investigate whether this could be a potentially faster means for 
obtaining equilibrated configurations. For the dense melts we studied we found,
however, that the fastest way to equilibrate the system was to use our standard
MD algorithm. The measured static
properties, as obtained in the MC simulations, were in good agreement with the 
measured static properties of the 
configurations produced with the Molecular Dynamics algorithm. Furthermore
the obtained energy distributions
were similar.
To check what influence the Nos\'e-Hoover thermostat has on the Newtonian dynamics
we also carried out some simulations in the microcanonical ensemble and compared
the results to the results of the simulations with Nos\'e-Hoover thermostat. 
Both methods lead to the same results, for example the velocity autocorrelation function
of the two simulations were identical (Figure 1). This means that the thermostat only
has a weak influence on the Newtonian dynamics although one is able to tune the 
temperature with it very effectively.

Starting
configurations were obtained using the method proposed by Kremer and Grest \cite{kremer,duenweg2}. Before subjected
to the Nos\'e-Hoover thermostat each thus generated configuration was propagated in the microcanonical
ensemble ($Q=\infty$). At the beginning of this step the velocities were rescaled several
times in order to come close to the desired temperature range. In the next equilibration
step the thermostat was switched on and the system was propagated until the mean square
displacement of the centers of mass of the polymer chains had reached several $R_g^2$, $R_g$ denoting
the radius of gyration.
At this time all measured correlators already had decayed to zero.

In order to speed up computational efficiency we applied a linked cell scheme combined with a
Verlet table \cite{link}. Because of the Nos\'e-Hoover thermostat it was not possible to
use a Velocity Verlet algorithm, instead we used a Heun algorithm \cite{gear} with a
time-step of $dt=0.002$.

All simulations in the NVT ensemble were performed using 95 polymer chains each consisting 
of 10 monomers. The volume was held constant at $V=1117.65 (\rho=0.85)$. Since the density 
was the same for all temperatures it was not necessary to generate starting
configurations at every temperature, as it was for the simulations in the NpT ensemble,
but one could use the ones generated and equilibrated at another temperature and equilibrate 
them again at the new temperature. Simulations were performed at temperatures 
$T=0.35, 0.38, 0.4, 0.45, 0.5, 0.6, 0.7, 1.0$ 
and $2.0$. For statistical reasons ten different configurations were simulated at each
temperature. The equilibration of a configuration at the lowest temperature required 
$30\times 10^6$ MD steps or almost two weeks of CPU time on a IBM Power PC 
for each configuration.

Since in the simulations of the NpT ensemble we wanted to keep the average pressure at
$p=1.0$ at all temperatures, the situation differed from the one of the simulation
of the NVT ensemble. In a first step we used a MD algorithm which also allowed 
for volume fluctuations of the system \cite{Hoover2,cicotti} to obtain the
average density of the system at a certain temperature. These runs lasted up to
$5 \times 10^6$ MD-steps. Afterwards, in a procedure analogous to
the one described above, we used the found density to generate starting configurations which we used for
NVT simulations. Note that we performed the simulations themselves at constant volume, but this
procedure made sure that the average pressure was constant (within five percent) 
at all temperatures. This was done because NVT simulations are computationally more efficient,
and also because we observed better stability for NVT than for NpT simulations.
At almost all temperatures we simulated 120 polymer chains again each
consisting of ten monomers. Furthermore at each temperature ten different configurations
were simulated and simulations were performed at temperatures 
$T=0.48, 0.5, 0.52, 0.55, 0.6, 0.65, 0.7, 1.0$ and $2.0$. 

\section{Static Properties}\label{secIII}
In this section we will discuss the static properties of the melts as a function of temperature.
The thermodynamic paths for our cooling processes in the NVT and NpT ensemble are shown in Fig. 2.
In the NVT ensemble we started our simulation at modestly high pressure at a high temperature. Upon
cooling the pressure decreases and becomes negative around $T=0.7$. This negative pressure
has consequences which we will discuss in detail when analysing the structure factor of the melt.
In the NpT ensemble we keep the pressure at ambient value and adjust the density upon cooling.

\subsection{Chain Conformations}
Let us now first look at the chain conformations upon cooling. The Hamiltonian we chose has no
intramolecular bond angle part and therefore there is no tendency of our chains to become
stiffer at lower temperatures. Consequently the size of the chains varies very little in
our whole temperature range ($R_g = 2.10$ at $T=0.35$, $R_g = 2.23$ at $T=2.0$ in the NVT ensemble
and $R_g = 2.09$ at $T=0.48$, $R_g = 2.23$ at $T=2.0$ in the NpT ensemble). In Fig. 3 we show the
behaviour of the structure factor of the chains in the NVT ensemble for the lowest simulation temperature. 
Also included is a Debye function \cite{DoiEdwards} calculated with the independently measured radius of gyration. 
The good agreement with the simulation data shows that the chains remain Gaussian on the
large scale over the whole temperature range. The stiffness parameter $C_N$ was in the range of
$1.51$ to $1.56$.

\subsection{Packing Behaviour}
The effect of the two competing length scales we introduced into our model can be nicely seen
looking at the monomer-monomer pair correlation function shown in Fig. 4. First of all we want
to note that the pair correlation function shows no long-range ordering even at the lowest
temperature we studied. The nearest neighbour peak, which is just a diffuse peak around $r=1.0$ at
high temperatures, splits in two upon cooling. The first of the peaks is due to the preferred
intramolecular distance or bond length $b=0.96$. The second peak is the preferred nearest neighbour
position in the minimum of the intermolecular Lennard-Jones interaction at $r_{min}=2^\frac{1}{6}$.

For the NVT ensemble
this real space behaviour transforms into the structure factor of the melt shown in Fig. 5. The
amorphous structure is here manifest in the amorphous halo around $q=6.9$
, which contains both intramolecular and intermolecular nearest neighbour
contributions. With decreasing temperature the short-range intermolecular order increases and
since this is the larger of the two length scales contributing to the amorphous halo, its position 
shifts to smaller $q$-values at first. At lower temperatures, however, this shift is reversed and
the peak moves to higher $q$-values as would be expected for thermal contraction of the sample.
In the same temperature range a small peak at very small $q$-values ($q\approx 1.7$) develops. 
Both effects result from a microvoid developing in the system because we are in a range of
negative pressure where the system would like to contract into a dense melt expelling the
free volume. 
The
microvoid, which contains up to around five percent of the simulation volume, can be identified 
both by visual inspection and numerical analysis. The position of the small $q$ peak is
related to the typical diameter of a microvoid. For the NVT simulations we therefore
have to keep in mind that at the lowest temperatures our system is no longer homogeneous
but contains a small amount of internal surfaces.

For the NpT simulations this effect is of course absent as can be seen in Fig 6. In this
case the structure factor shows the behaviour also seen experimentally with the amorphous
halo moving to larger $q$-values due to the increased density at lower temperatures.

\section{Dynamic Properties}\label{secIV}

In this section we will look at the temperature dependence of the largest relaxation time
in the melt. For simple glass forming liquids this is called the $\alpha$-relaxation time.
This is the timescale at which a particle breaks free of the cage of its nearest neighbours
and large scale structural relaxation becomes possible. For polymers this then also is
the timescale on which local conformational rearrangements start to occur. The largest
relaxation time in polymers, however, is the time for the overall renewal of the chain
conformation, which is a factor of $N^2$ ($N$ being the number of monomers in a polymer
chain) larger for chains following Rouse dynamics \cite{Rouse}
and a factor of $N^3$ for larger chains where reptation effects have to be taken into account
\cite{DoiEdwards}. The temperature dependence of this longest relaxation time is 
determined by the temperature dependence of the prefactor in these scaling laws, 
which is the timescale for local conformational changes, which, as discussed, is enslaved
to the $\alpha$-process of the structural relaxation.

\subsection{Structural Relaxation}
We will discuss the structural relaxation in terms of the incoherent intermediate dynamic
structure factor
\begin{equation}
F_q(t) = \left<\frac{1}{M}\sum_{i=1}^{M} e^{i {\bf q}({\bf r}_i(t)-{\bf r}_i(0))}\right> ,
\end{equation}
where $M$ stands for the total number of monomers in the melt. As can be seen in Fig. 7, which shows $F_q(t)$ at
the peak position of the static structure factor, the intermediate dynamic structure factor starts
to exhibit a two step relaxation process when lowering the temperature, the so called $\beta$- and
$\alpha$-processes. In this paper we focus on the behaviour of the long-time $\alpha$-process
and leave a detailed analysis of the $\beta$-relaxation to a forthcoming publication. 

Comparing the behaviour of the NVT simulations (Fig. 7a) and the NpT simulations (Fig 7b) we 
see that the slowing down of the structural relaxation and the development of the two step
process occur for higher temperatures in the NpT ensemble.

The behaviour of $F_q(t)$ at the first minimum of the static structure factor is qualitatively the same,
with the plateau region occurring at a smaller value of $F_q(t)$. We empirically define an $\alpha$-
relaxation timescale by the requirement:
\begin{equation}
F_q(\tau_\alpha)=0.3 \: .
\end{equation}
In the undercooled liquid close to the mode coupling critical temperature the time-temperature
superposition principle is expected to hold. Figures 8a and 8b show that indeed we find a superposition of
the $\alpha$-relaxation behaviour for the NVT simulation in the region $0.35 < T < 0.45$
and for the NpT simulations it the range $0.48 < T < 0.6$.

One generally analyzes the $\alpha$-relaxation behaviour also by fitting the empirical KWW function
to the data:  
\begin{equation}
f(t)=A e^{-(\frac{t}{\tau})^\beta}
\end{equation}
In the temperature range where we found the time-temperature superposition principle to hold we find:
\begin{equation}
\beta = 0.56 \pm 0.04 (NVT); \beta = 0.7 \pm 0.08 (NpT)
\end{equation}
The error bars are mostly due to the effect that one can change $\beta$ by almost 15 \% by changing
the time interval over which one tries to fit the data. When one tries to fit the behaviour at higher
temperatures the resulting values for $\beta$ increase approaching unity at high temperatures.

For the $\alpha$-relaxation timescale the mode coupling theory of the glass transition predicts a power
law divergence:
\begin{equation}
\tau_\alpha \propto (T-T_c)^{-\gamma}
\end{equation}
Fig. 9 shows that we indeed observe this behaviour with $T_c$ and $\gamma$ depending on the thermodynamic
ensemble. For the NVT simulations we obtain $T_c = 0.32 \pm 0.01$ and $\gamma = 2.3 \pm 0.2$  and for 
the NpT simulations we have $T_c =0.45 \pm 0.01$ and $\gamma = 1.95 \pm 0.15$. For our model we
therefore equilibrated our system to temperatures within 10 \% of $T_c$. Note that near $T_c$ in the
NpT ensemble the density is much higher than for our choice of density in the NVT ensemble and thus
the large difference of $T_c$ is not unexpected.

However it should be
mentioned
that it is in principle also possible to fit our data with the well known Vogel-Fulcher-Tammann (VFT)
equation:
\begin{equation}
\tau_\alpha = \tau_0 \exp({\frac{E}{T-T_0}})
\end{equation}
We obtain $T_0 = 0.215 \pm 0.02$ and $E=1.1 \pm 0.1$ for the NVT-simulations and 
$T_0 = 0.34 \pm 0.02$ and $E=0.93 \pm 0.1$ for the NpT-simulations. Close to the mode coupling critical temperature
the VFT curve is well approximated by a power law divergence with exactly the same critical
temperature and exponent as obtained from an independent mode coupling fit. Very close to $T_c$ the VFT 
curve flattens in comparison with the ideal mode coupling fit. This is again in accord with what would
be predicted by an extended mode coupling analysis \cite{baschnag1} taking into account structural
decay via activated processes. It is also a behaviour typically seen in experiment \cite{Li2}
and simulations \cite{kaemmerer}.

\subsection{Polymer Self-Diffusion}
The overall conformational relaxation of polymer molecules can be conveniently analyzed by
looking at their self-diffusion behaviour \cite{paul_jcp}. For this purpose one can look at:

\begin{equation}
g_1(t) = \left<({\bf r}_{\frac{N}{2}}^j(t)-{\bf r}_{\frac{N}{2}}^j(0))^2\right>,
\end{equation}
which describes the mean square displacement of the inner monomers ($j$ labels different
polymer chains).
The analogous quantity in the center of mass reference frame of chain $j$ (${\bf r}_{cm}^j(t)$ being
the position of the center of mass of polymer $j$ at time $t$) is:
\begin{equation}
g_2(t) = \left<({\bf r}_{\frac{N}{2}}^j(t)-{\bf r}_{cm}^j(t)-{\bf r}_{\frac{N}{2}}^j(0)+{\bf r}_{cm}^j(0))^2\right>.
\end{equation}\noindent
The mean square displacement of the center of mass itself is:
\begin{equation}
g_3(t) = \left<({\bf r}_{cm}^j(t)-{\bf r}_{cm}^j(0))^2\right>.
\end{equation}
\noindent
And finally the mean square displacement of monomers at the free ends of the chains 
and its analogous quantity in the center of mass reference frame are defined as:
\begin{equation}
g_4(t) = \left<({\bf r}_{end}^j(t)-{\bf r}_{end}^j(0))^2\right>
\end{equation}
\begin{equation}
g_5(t) = \left<({\bf r}_{end}^j(t)-{\bf r}_{cm}^j(t)-{\bf r}_{end}^j(0)+{\bf r}_{cm}^j(0))^2\right>.
\end{equation}
\noindent
Fig. 11a shows $g_1$ to $g_5$ measured at $T=1.0$ and Fig. 11b the same quantities at $T=0.35$.
For $g_1(t)$ one can distinguish several regimes. For short times $t<0.1$ one observes 
a ballistic regime which is followed by a subdiffusive regime and finally a free diffusion 
regime. Such a behaviour is typical
for polymer systems and is predicted by many theories, e.g. the Rouse model.
As can be seen in Fig. 11b,  the situation is a little bit different at lower temperatures.
The ballistic regime is now followed by a plateau like regime which precedes the subdiffusive one.
Such a plateau regime is typical for glass formers and a sign of the onset of the structural arrest 
of the system. The height of the plateau is closely related to the size of the cage a particle is 
trapped in. 
Another difference to high temperatures is that  the subdiffusive regime stretches 
out far more in time 
and, therefore, the free diffusion limit is reached only after long simulation times. 

\noindent
The subdiffusive regime can be fitted using:
\begin{equation}
g_1(t)= \sigma^2 (W_1\:t)^{x_1}.
\end{equation}

\noindent
We obtain $x_1 = 0.62 \pm 0.02$ for all simulated temperatures. $g_4(t)$ behaves
similar to $g_1(t)$. Here the diffusive regime is preceded by
a subdiffusive one as well, which can be fitted by:
\begin{equation}
g_4(t)= \sigma^2 (W_4\:t)^{x_4}.
\end{equation}

\noindent
Again the exponent is approximately the same at all temperatures. We
find $x_4 = 0.67 \pm 0.03$.

If our model chains would exactly follow the Rouse predictions the local monomer mobilities $W_1$ and $W_4$ should
be equal and $x_1$ and $x_4$ should be equal to $0.5$. It is however a general finding
from simulations, that for chains as short as ours one generally observes a smeared out 
crossover behaviour \cite{paul2} from short time to long time diffusion instead of the predicted
Rouse exponent.
At later times $g_1(t)$, $g_3(t)$ and $g_4(t)$ all show the expected
simple diffusive behaviour:
\begin{equation}
g_i(t)= 6\: D \: t,
\end{equation}
where the self diffusion constant is the same for all $i$. At lower temperatures,
especially for $g_1(t)$ and $g_4(t)$, it is very hard to distinguish this regime
from the preceding subdiffusive one. 

\noindent
In Figure 12 $D(T)$, $W_1(T)$ and $W_4(T)$ are plotted against the temperature. As one
sees for lower temperatures all quantities follow a power law behaviour:

\begin{eqnarray}
D & \propto & (T-T_c)^\gamma \\
W_i & \propto & (T-T_c)^\gamma \: , \: i=1, 4
\end{eqnarray}
The critical temperature and the exponent are, within range of error, the same as those for the 
$\alpha$-relaxation
timescale in the respective ensemble. 
This shows the coupling of the conformational
relaxation and diffusion of the polymer chains to the local structural $\alpha$-relaxation
in the melt as discussed in the beginning of this section. Note that it is again possible to fit 
the data with the VFT-equation using the same $T_0$ and $E$ (within range of error) as obtained
when fitting the $\alpha$-relaxation times.

\section{Conclusions}\label{secV}
In this paper we have presented a Molecular Dynamics simulation of the thermal glass transition
in  dense polymer melts. We have studied this transition at constant density as well as 
constant pressure. Our model is a coarse-grained bead-spring model with nonlinear springs 
connecting monomers along a chain and Lennard-Jones interactions between all monomers. In order to
introduce packing frustration into the model we chose
incompatible length scales for intra- and intermolecular nearest neighbour distances. All our
results were obtained on well equilibrated samples.

We showed that the static structure factor of our chains of length $N=10$ could be well described
by a Debye function at all temperatures. The size of the chains is mostly temperature independent
as we introduced no temperature dependent stiffness into the model. The two incompatible length
scales in the Hamiltonian can be seen in a split of the first neighbour peak of the
monomer-monomer pair correlation function at low temperatures. For the constant volume simulation
the density we chose led to negative pressure for $T<0.7$. This instability led to the buildup
of microvoids taking up approximately {5 \%} of the simulation volume at low temperatures. 
The observed negative pressure is an indication that the void formation process is not fully
completed on the timescale of the simulation.

In this work we analysed the glass transition in terms of the $\alpha$-relaxation process.
The divergence of the $\alpha$-relaxation timescale could be very well described by the power
law behaviour predicted by MCT. Critical temperatures differ substantially and power law exponents differ 
slightly between
cooling at constant volume and cooling at constant pressure. The divergence of the $\alpha$-relaxation
timescale also leads to a divergence of the largest relaxation time in the system
which is the Rouse time for the overall renewal of the chain conformations. This
divergence can be observed looking at the mobility of monomers on intermediate length scales
(as measured by the rate constants $W_1$ and $W_4$) and the center of mass self-diffusion
coefficient of the chains. All these quantities follow power law singularities with values for the
critical temperatures and exponents in nice agreement with the behaviour of the $\alpha$-relaxation
timescale. The divergence could be equally well described by a VFT fit to the data and it could
be shown that close to $T_c$ the MCT power law singularity is a tangent approximation of the
VFT curve.

A detailed analysis of the $\beta$-scaling regime predicted by mode coupling theory of the glass
transition will be presented in a separate publication.

\section{Acknowledgements}
We would like to thank J. Baschnagel and W. Kob for helpful discussions, and A. Kopf for 
supplying his MD code which was modified for the present study. Support
by the Sonderforschungsbereich SFB 262 and generous grants of computer time from the
computing center of the University of Mainz and the HLRZ J\"ulich are gratefully acknowledged.

\newpage

\begin{figure}[]
\centering
\includegraphics[width=110mm,height=90mm]{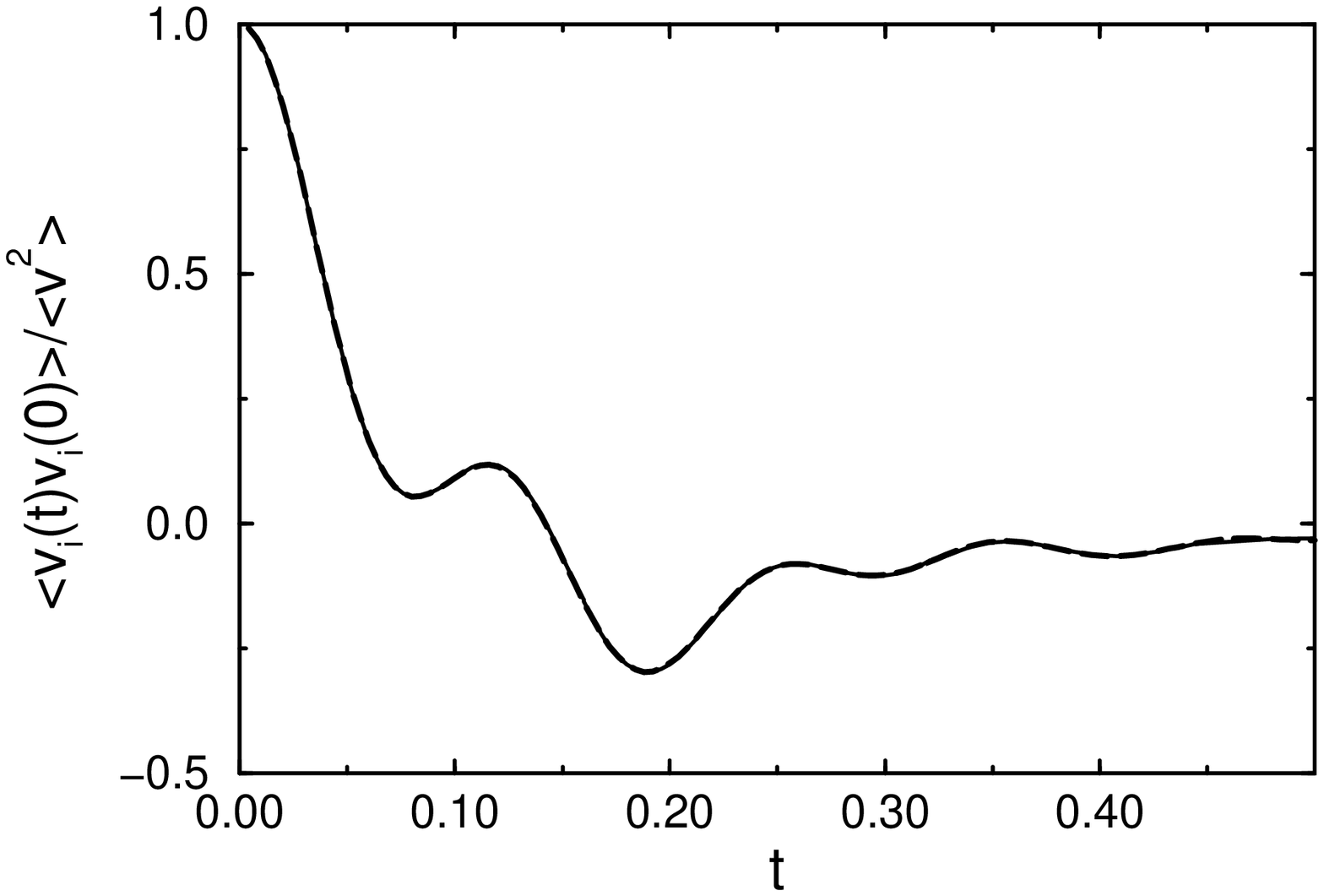}
\caption{Velocity autocorrelation function with and without (dashed line)
Nos\'e-Hoover thermostat. The two lines are practically identical 
meaning that the thermostat has only a weak influence on the Newtonian
dynamics. Note that because we were using Lennard-Jones units all 
quantities shown are dimensionless.}
\end{figure}

\begin{figure}[]
\centering
\includegraphics[width=110mm,height=90mm]{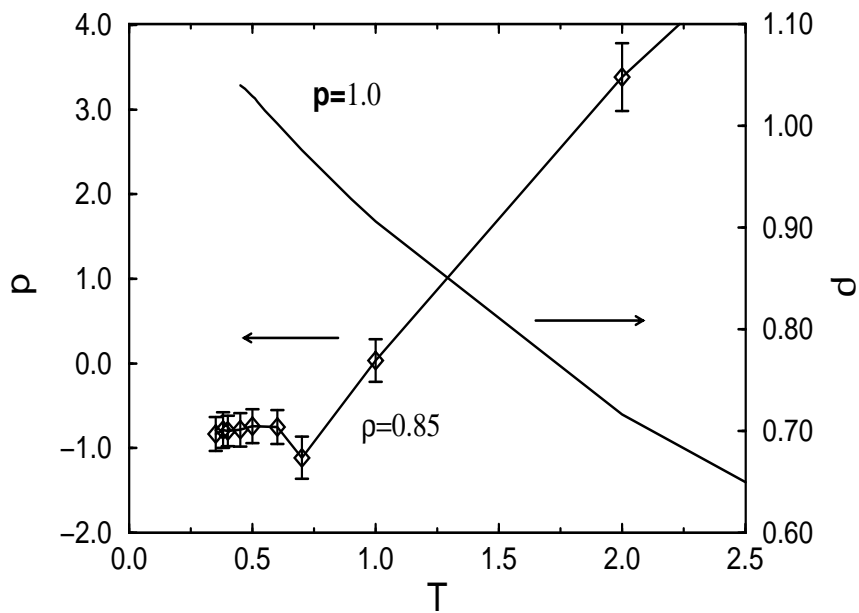}
\caption{Thermodynamic paths for the cooling experiments we performed
in the simulation.}
\end{figure}

\begin{figure}[]
\centering
\includegraphics[width=110mm,height=90mm]{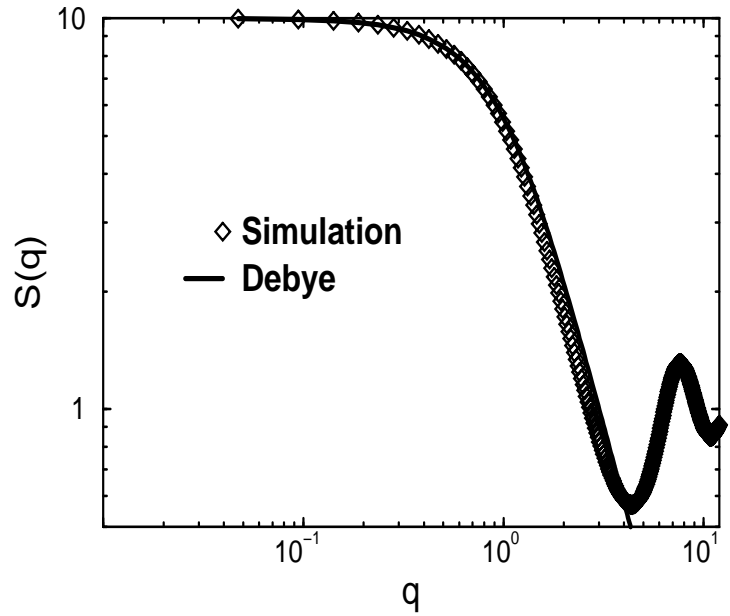}
\caption{Structure factor of the chains for the lowest simulation temperature. Also 
shown is the Debye function corresponding to the independently measured
radius of gyration.}
\end{figure}

\begin{figure}[]
\centering
\includegraphics[width=110mm,height=90mm]{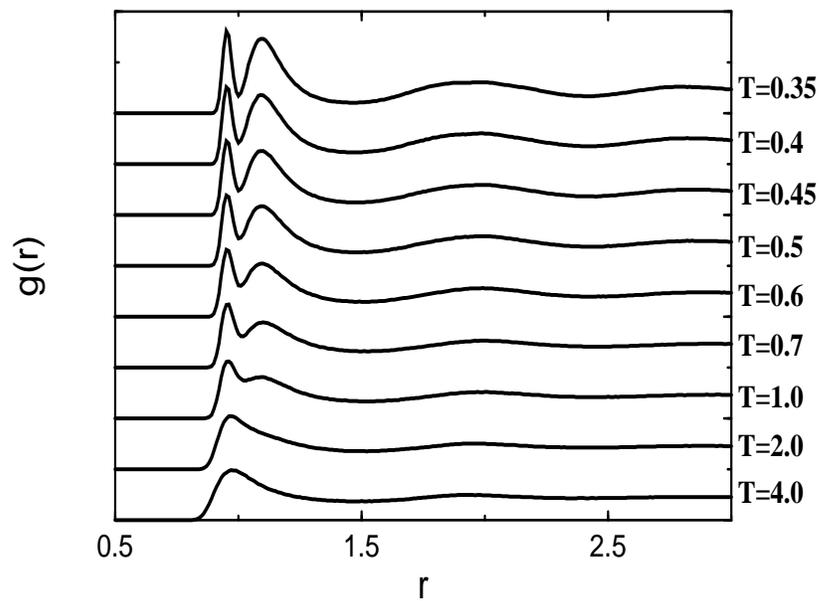}
\caption{Pair correlation function for the NVT simulations and the range
of indicated temperatures. Curves for lower temperatures are shifted
upward.}
\end{figure}

\begin{figure}[]
\centering
\includegraphics[width=110mm,height=90mm]{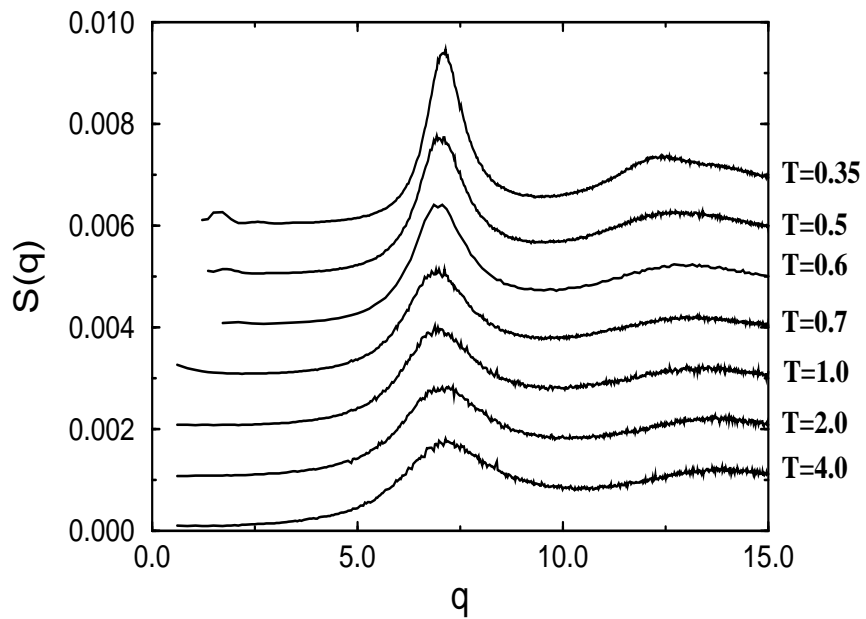}
\caption{Structure factor of the melt for a set of temperatures in the
NVT simulations. Note the appearance of a small peak at
low q-values around T=0.6 which is due to the emergence of a
micro void in the system.}
\end{figure}

\begin{figure}[]
\centering
\includegraphics[width=110mm,height=90mm]{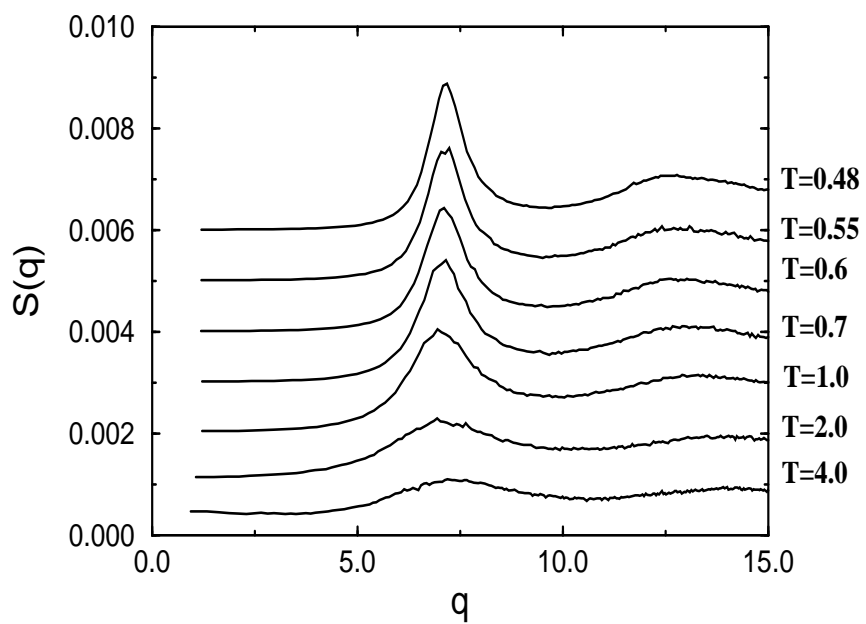}
\caption{Same as Figure 5 for the NpT simulations.}
\end{figure}

\begin{figure}[]
\centering
\includegraphics[width=110mm,height=90mm]{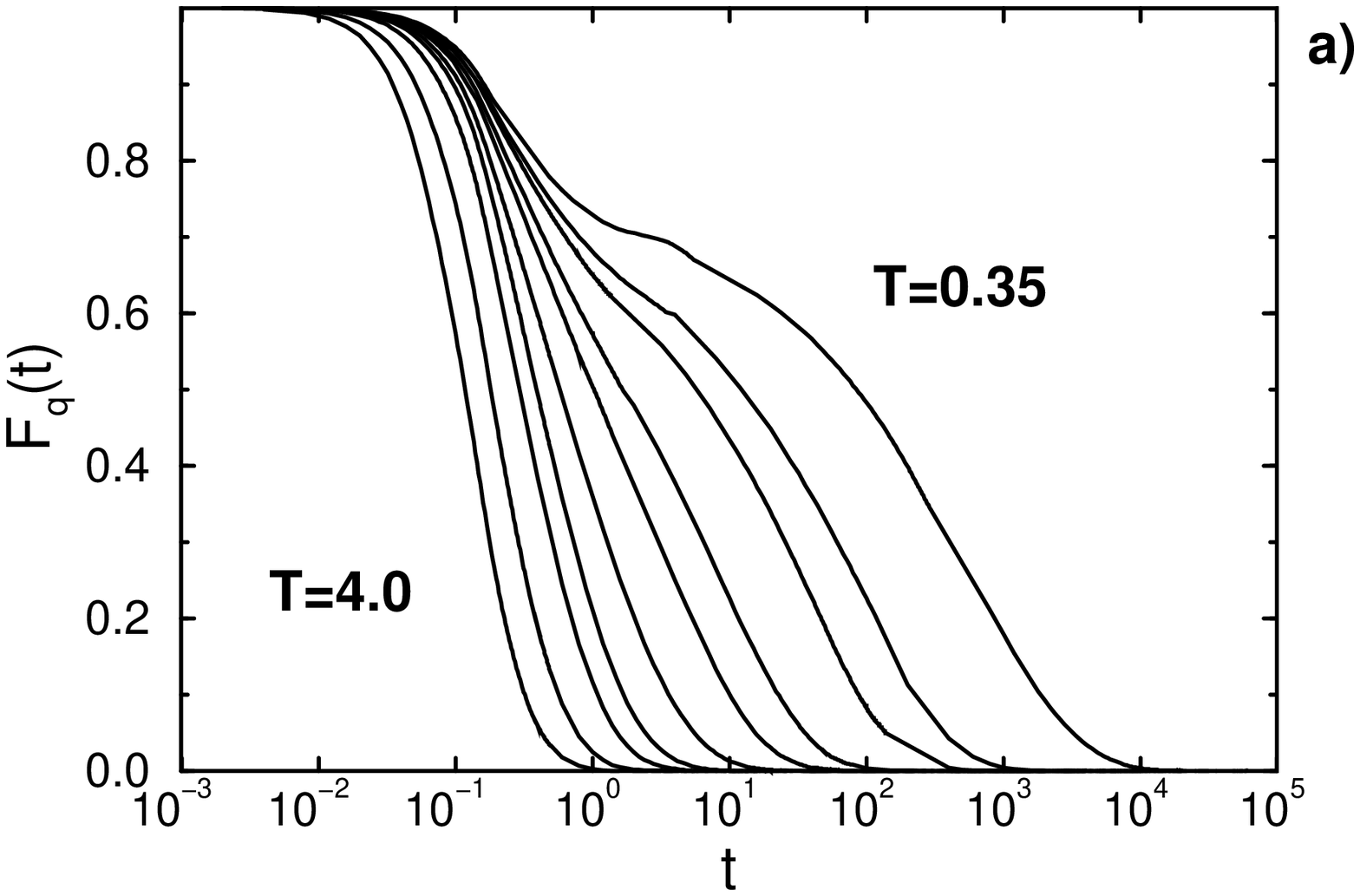}
\includegraphics[width=110mm,height=90mm]{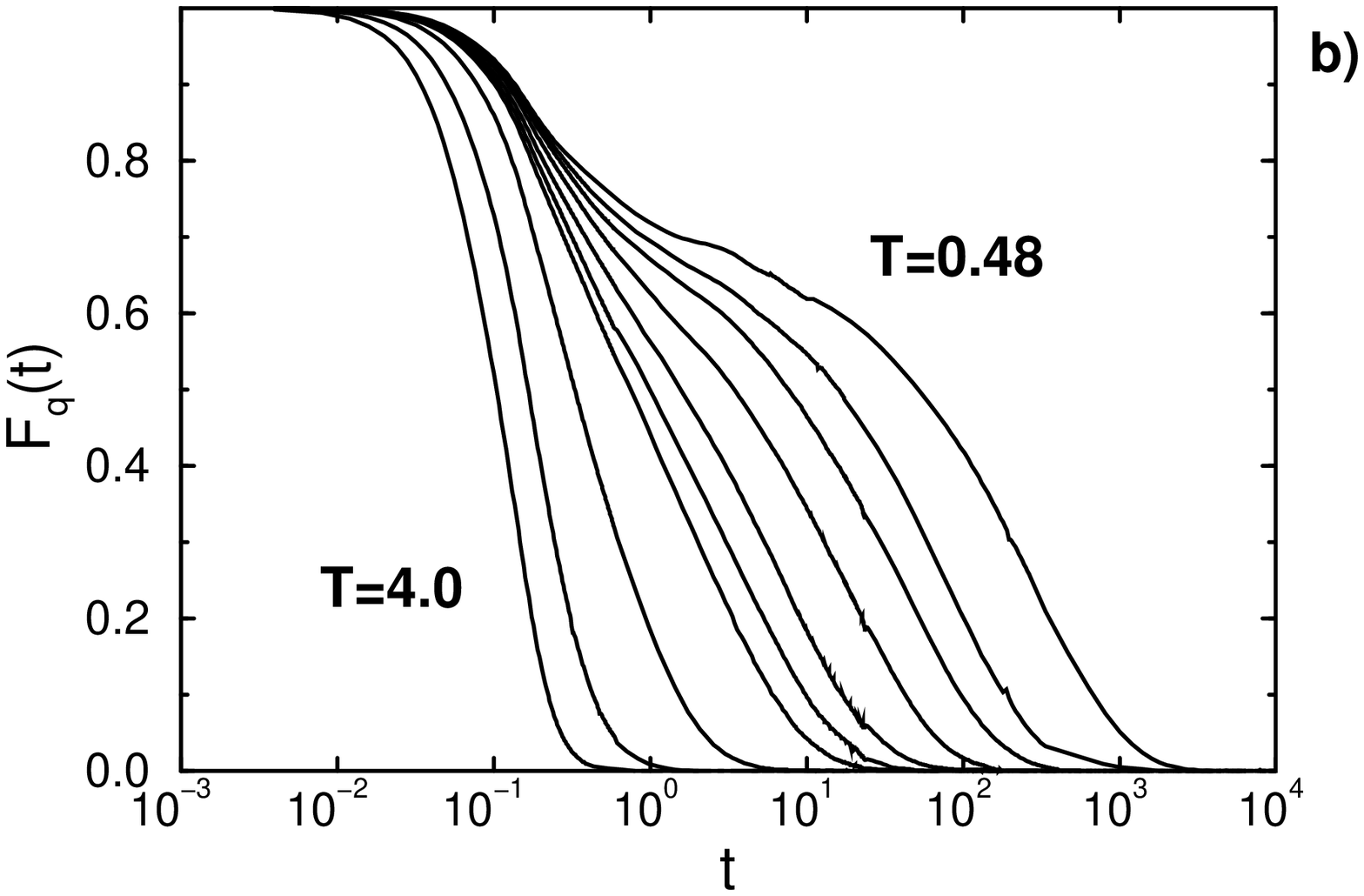}
\caption{Intermediate dynamic structure factor at the first maximum
of the static structure factor for the NVT simulations (a) and the same for 
the NpT simulations (b).}
\end{figure}

\begin{figure}[]
\centering
\includegraphics[width=110mm,height=90mm]{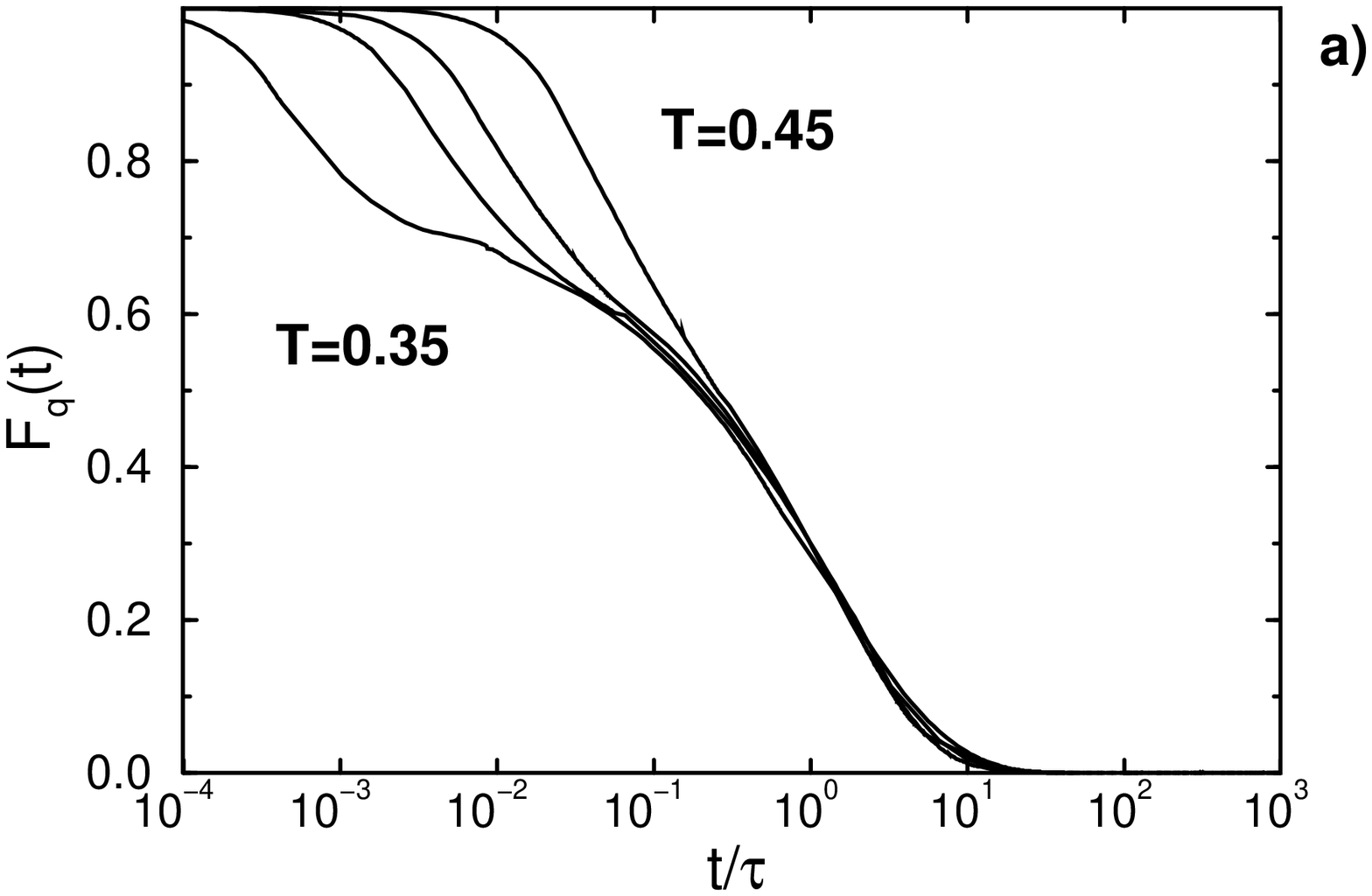}
\includegraphics[width=110mm,height=90mm]{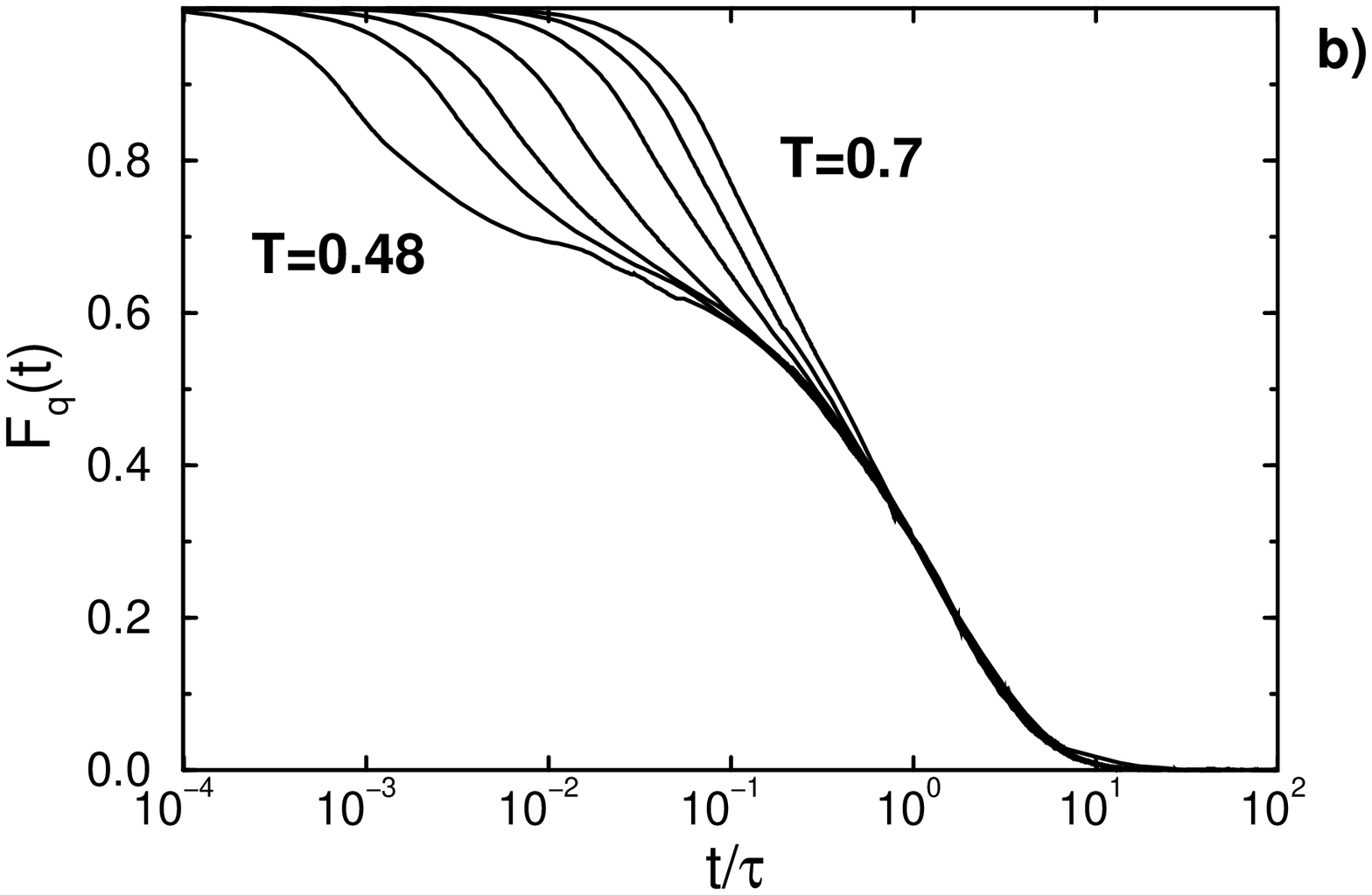}
\caption{Same as Figure 7 but with times scaled by the $\alpha$-relaxation
timescale for the range of temperatures where the time-temperature
superposition holds.  (a) NVT ensemble, (b) NpT ensemble.}
\end{figure}

\begin{figure}[]
\centering
\includegraphics[width=110mm,height=90mm]{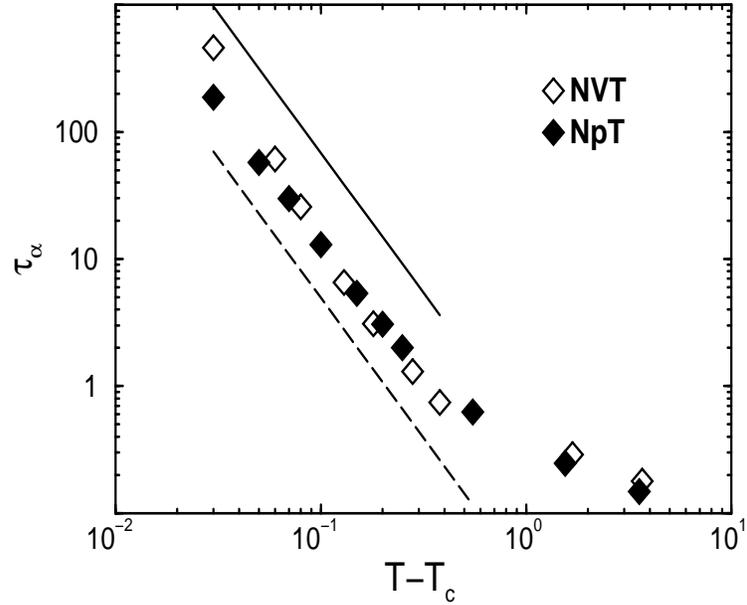}
\caption{Critical behaviour of the $\alpha$-relaxation timescale close
to the respective critical temperatures, $T_c=0.32$ in the NVT ensemble
and $T_c=0.45$ in the NpT ensemble, plotted against $T-T_c$
. Open diamonds are for the NVT ensemble
and closed diamonds are for the NpT ensemble.}
\end{figure}

\begin{figure}[]
\centering
\includegraphics[width=110mm,height=90mm]{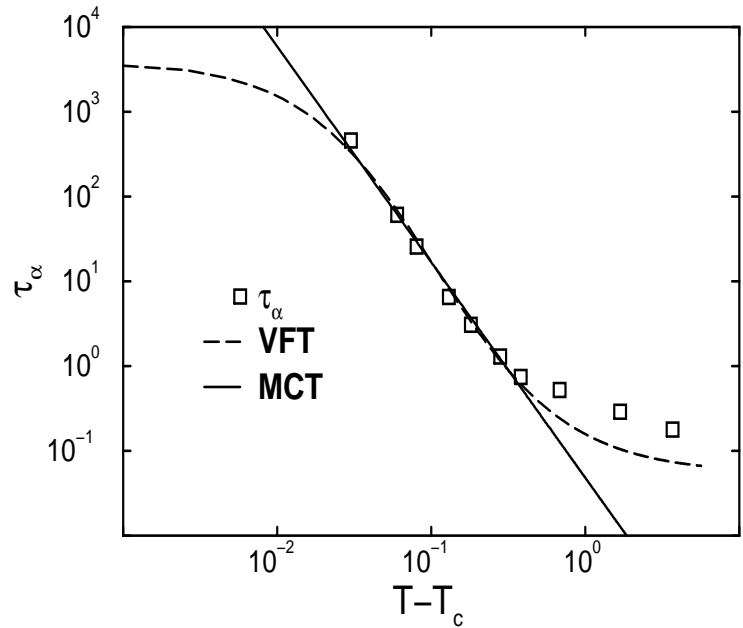}
\caption{$\tau_\alpha$ as measured in the NVT ensemble. Also shown are
best fits with the VFT-equation and the predictions of MCT. The temperature
range is shifted by $T_c=0.32$ to show the similarity of the predictions 
of the two equations for our data close to $T_c$. Note that very similar plots
can be obtained for the quantities discussed in Fig. 12.}
\end{figure}

\begin{figure}[]
\centering
\includegraphics[width=110mm,height=90mm]{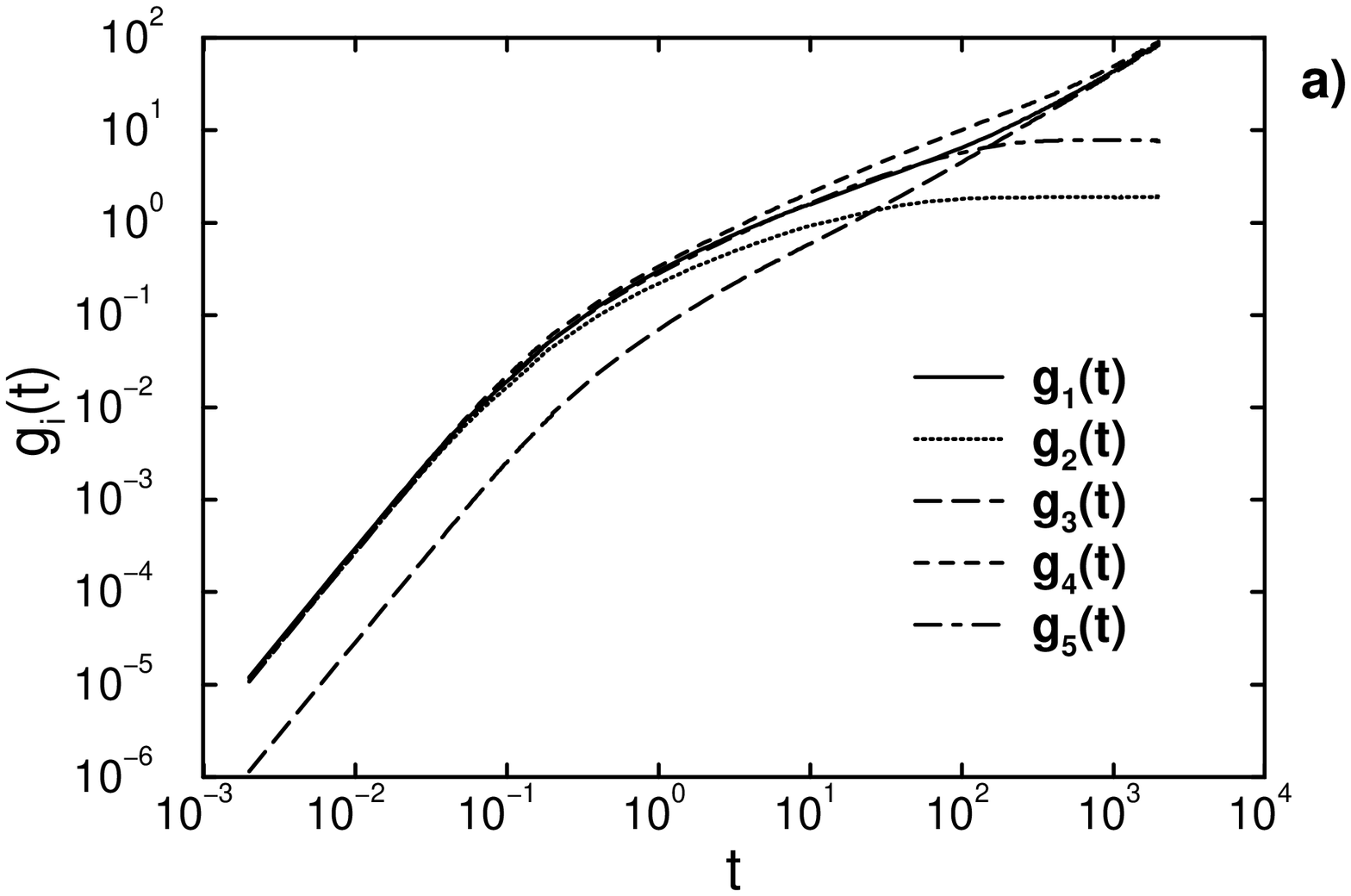}
\includegraphics[width=110mm,height=90mm]{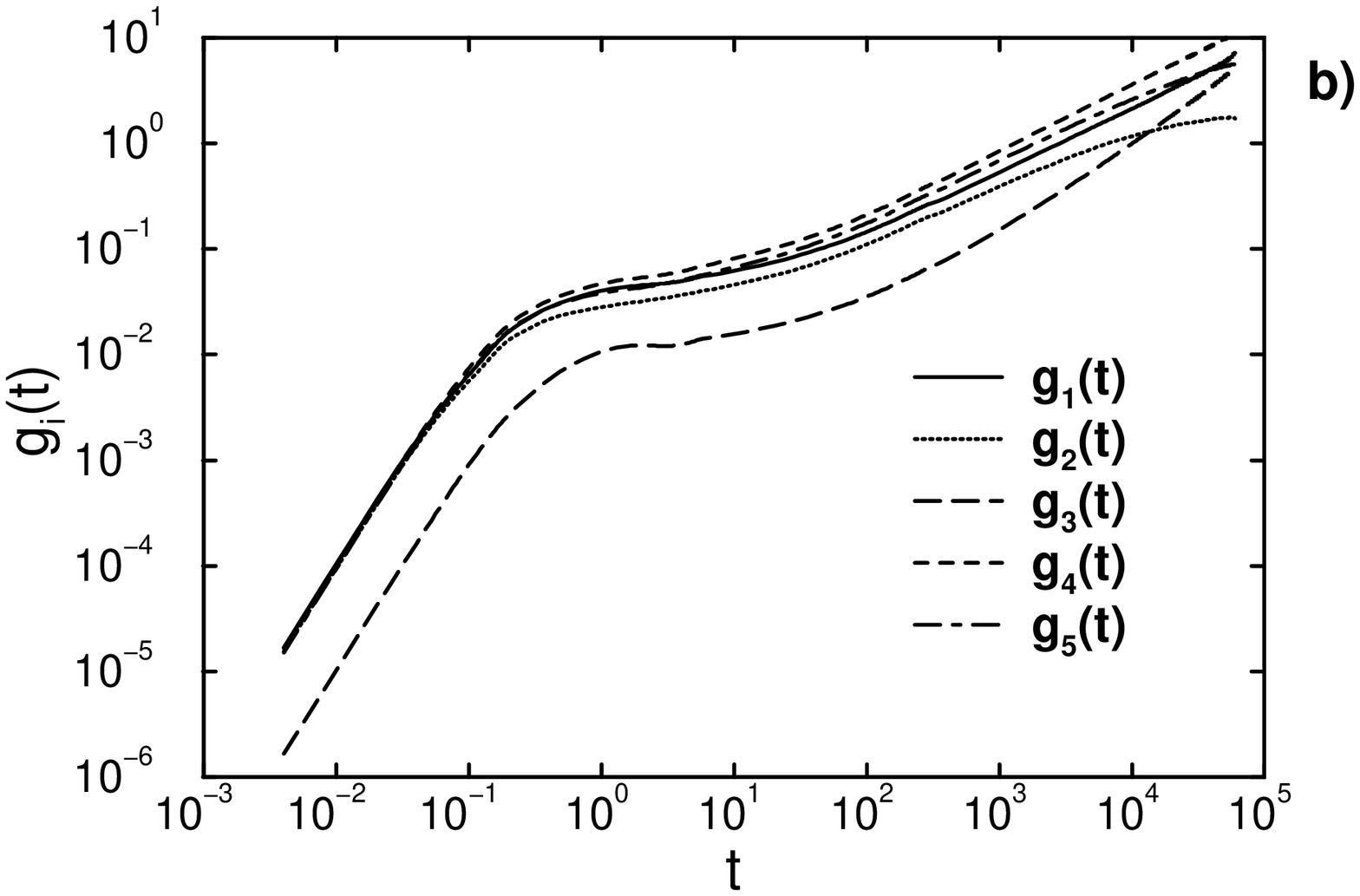}
\caption{Mean square displacements as a function of time for $T=1.0$ (a)
 and $T=0.35$ (b). $g_3$ describes the center of 
mass of the chain, $g_1$ and $g_4$ are center and end monomers respectively
and $g_2$ and $g_5$ are the corresponding displacements in the center of
mass reference frame. }
\end{figure}

\begin{figure}[]
\centering
\includegraphics[width=110mm,height=90mm]{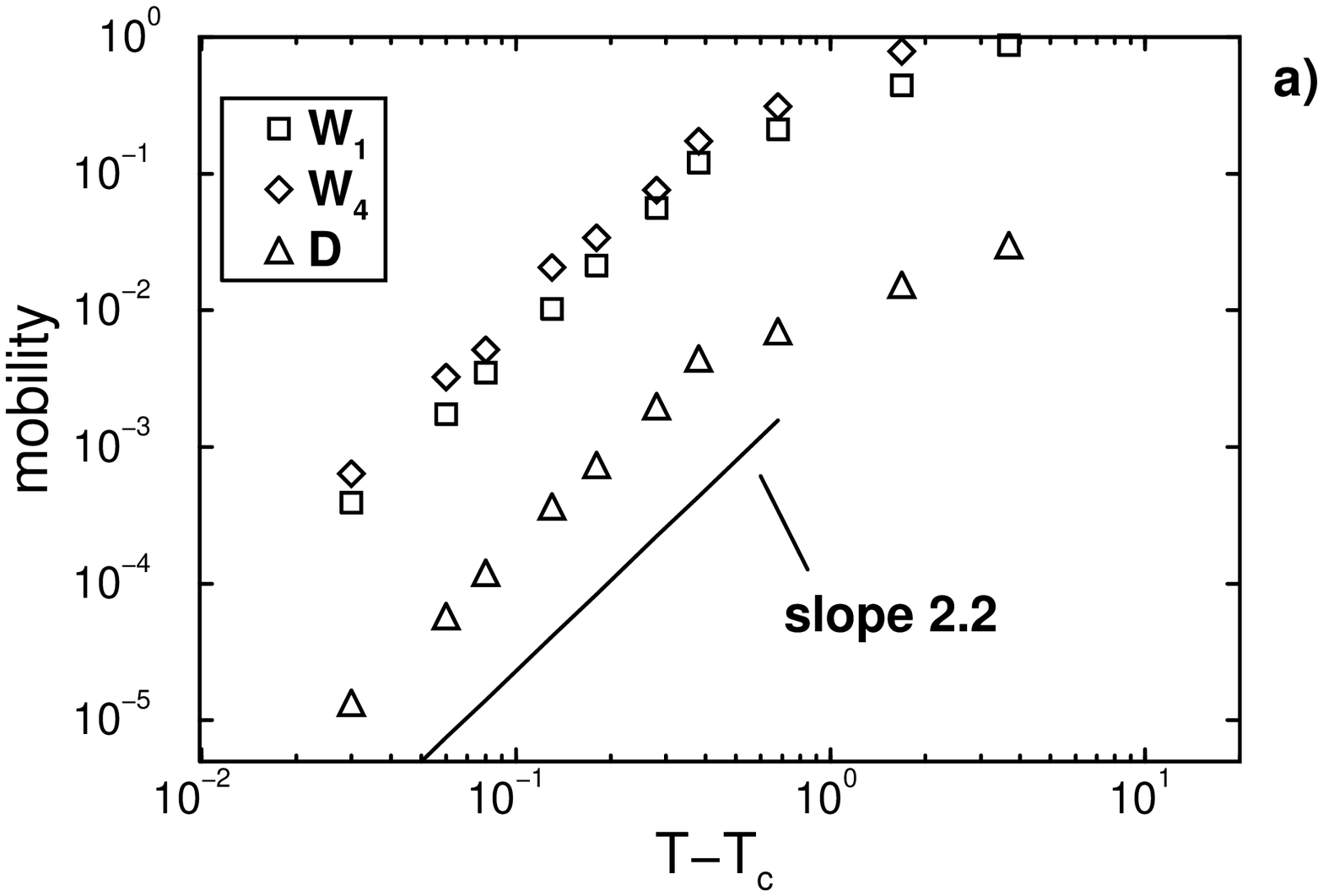}
\includegraphics[width=110mm,height=90mm]{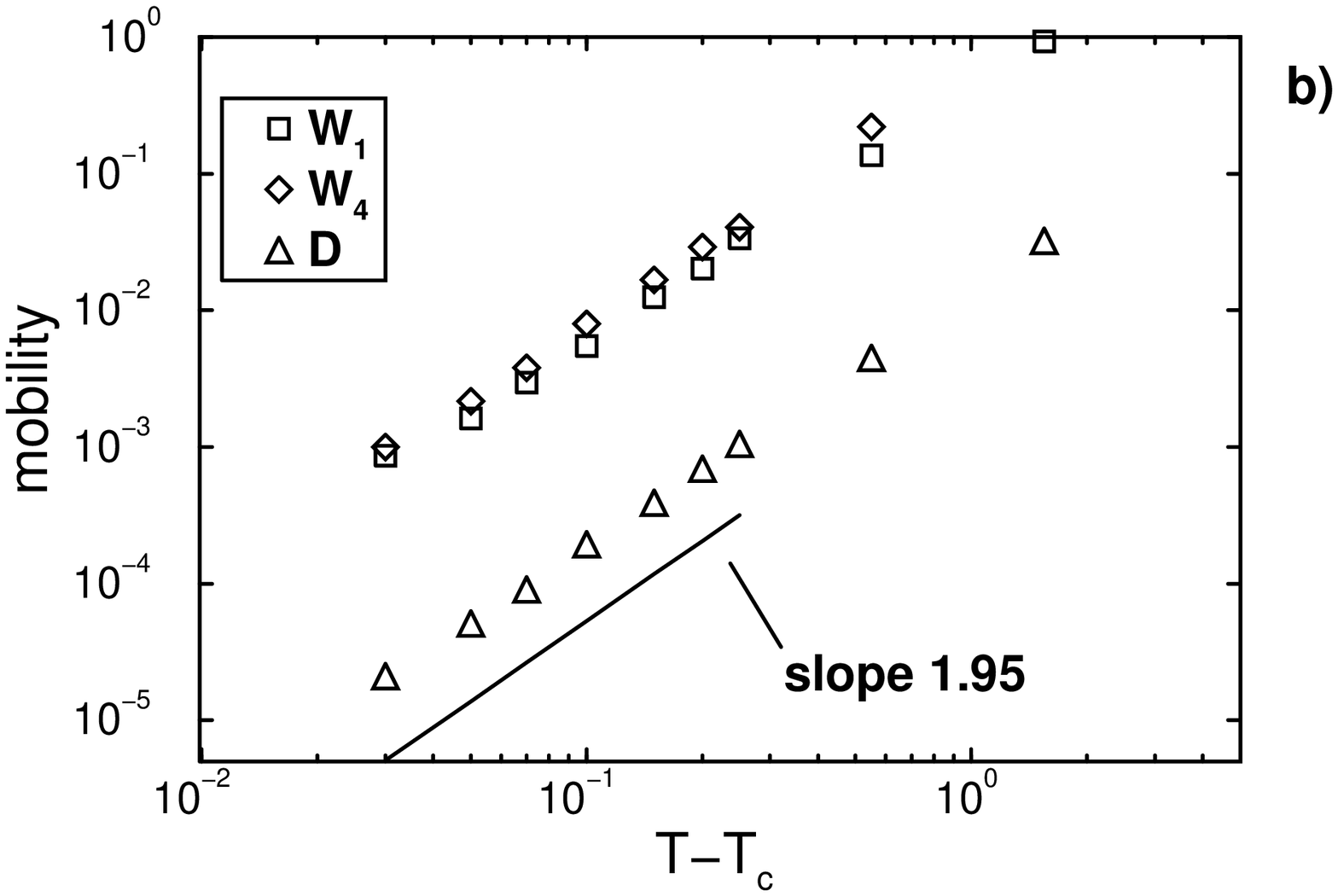}
\caption{Critical behaviour of the self-diffusion coefficient and
the rate constants for the NVT (a) and NpT simulations (b).}
\end{figure}


\begin{thebibliography}{99}

\bibitem{jaeckle} J.\ J\"ackle, Rep.\ Prog.\ Phys.\ {\bf 49}, 171 (1986)
\bibitem{goetze1} W.\ G\"otze and L.\ Sj\"ogren, J.\ Phys.\ C.: Solid State Phys.\ {\bf 20}, 879 (1987)
\bibitem{technik} {\it Material Science and Technology}, Vol 9, J. Zarzycki ed. (VCH, Weinheim, 1991)
\bibitem{McKenna} G.\ B.\ McKenna, in {\it Comprehensive Polymer Science}, Vol 2, edited by C.\ Booth and
C.\ Price (Pergamon, New York, 1989) 
\bibitem{DiMarzio} J.\ H.\ Gibbs and E.\ A.\ DiMarzio, J.\ Chem.\ Phys.\ {\bf 28} (3), 373 (1958)




\bibitem{sjlander} U.\ Bengtzelius, W.\ Goetze and A.\ Sj\"olander, J.\ Phys.\ C {\bf 17}, 59115 (1984)
\bibitem{goetze2} W.\ G\"otze and L.\ Sj\"ogren, Rep.\ Prog.\ Phys.\ {\bf 55}, 241 (1992)
\bibitem{goetze3} W.\ G\"otze and L.\ Sj\"ogren, Transport Theory and Statistical Physics {\bf 24}, 801 (1995)
\bibitem{leutheusser} E.\ Leutheusser, Phys.\ Rev.\ A {\bf 29} (5), 2765 (1994)


\bibitem{Heraklion} {\it Proceedings of 1st International Discussion Meeting on Relaxations in Complex
Systems },  J.\ of Non-Cryst.\ Sol. {\bf 131}-{\bf 133} (1991)
\bibitem{Alicante} {\it Proceedings of 2nd International Discussion Meeting on Relaxations in Complex
Systems }, J.\ of Non-Cryst.\ Sol. {\bf 172}-{\bf 174} (1994)
\bibitem{Vigo} {\it Proceedings of 3rd International Discussion Meeting on Relaxations in Complex
Systems }, J.\ of Non-Cryst.\ Sol.\ (to be published)

\bibitem{puce} P.\ N.\ Pusey and W.\ van Megen, Nature {\bf 320}, 340 (1986); Phys.\ Rev.\ Lett.\ {\bf 59}, 2083 (1987)
\bibitem{vanMegen} W.\ van Megen and S.\ M.\ Underwood, Phys.\ Rev.\ E {\bf 47} (1), 248 (1993)
\bibitem{mezei} F.\ Mezei, W.\ Knaak and B.\ Farago, Phys.\ Rev.\ Lett.\ {\bf 58}, 571 (1987)
\bibitem{roessler} E.\ Roessler and H.\ Sillescu, in {\it Material Science and Technology}, Vol 9, J. Zarzycki ed. (VCH, Weinheim, 1991)
\bibitem{Frick} B.\ Frick, B.\ Farago and D.\ Richter, Phys.\ Rev.\ Lett.\ {\bf 64} (24), 2921 (1990)


\bibitem{kob2} W.\ Kob and H.\ C.\ Andersen, Phys.\ Rev.\ E {\bf 51}, 4626 (1995)
\bibitem{kob3} W.\ Kob and H.\ C.\ Andersen, Phys.\ Rev.\ E {\bf 52}, 4134 (1995)
 

\bibitem{Nauroth} M.\ Nauroth and W.\ Kob , Phys.\ Rev.\ E {\bf 55}, 657 (1997)

\bibitem{paul2} W.\ Paul and J.\ Baschnagel, in {\it Monte Carlo and Molecular Dynamics Simulations in Polymer Sciences},
ed. by K.\ Binder, (Oxford University Press, Oxford, 1995)


\bibitem{baschnag1} J.\ Baschnagel and M.\ Fuchs, J.\ Phys.\  Condens.\ Matter {\bf 7}, 6761 (1995)
\bibitem{baschnag2} J.\ Baschnagel, Phys.\ Rev.\ B {\bf 49}, 135 (1994)
\bibitem{Wolf1} M.\ Wolfgardt, J.\ Baschnagel, W.\ Paul and K.\ Binder,
Phys.\ Rev.\ E {\bf 54} (2), 1535 (1996)
\bibitem{Wolf2} M.\ Wolfgardt and K.\ Binder, Macromol. Theory Simul. {\bf 5},
699 (1996)
\bibitem{Wolf3} J.\ Baschnagel, M.\ Wolfgardt, W.\ Paul and K.\ Binder,
J.\ Res.\ Natl.\ Inst.\ Stand.\ Technol.\ {\bf 102}, 159 (1997)
\bibitem{okun} K.\ Okun, M.\ Wolfgardt, J.\ Baschnagel and K.\ Binder,
Macromolecules {\bf 30} (10), 3075 (1997)
\bibitem{paul1} W.\ Paul, K.\ Binder, K.\ Kremer and D.\ W.\ Heermann, Macromolecules {\bf 24}, 6332 (1991);
W.\ Paul and N.\ Pistoor, Macromolecules {\bf 27}, 1249 (1994); 
V.\ Tries, J.\ Baschnagel, W.\ Paul and K.\ Binder, J.\ Chem.\ Phys.\ {\bf 106}, 738 (1997)
 

\bibitem{brown} D.\ Brown and J.\ H.\ R.\ Clarke, J.\ Chem.\ Phys.\ {\bf 84}, 2858 (1986)

\bibitem{roe1} D.\ Rigby and R.-J.\ Roe, J.\ Chem.\ Phys.\ {\bf 87} (12), 7285 (1987)
\bibitem{troe} H.\ Takeuchi, R.-J.\ Roe and J.\ E.\ Mark, J.\ Chem.\ Phys.\ {\bf 93}, 9042 (1990)
\bibitem{roe2} R.-J.\ Roe, J.\ Chem.\ Phys.\ {\bf 100} (2), 1610 (1994)

\bibitem{kremer} K.\ Kremer and G.\ S.\ Grest, J.\ Chem.\ Phys.\ {\bf 92} (8), 5057 (1990)
\bibitem{Koppelmann} J.\ Koppelmann, in {\em Proceedings of the fourth international 
congress on rheology}, Vol.\ 3, ed. by E.\ H.\ Lee and A.\ L.\ Copley (Wiley, New York, 1965)
\bibitem{duenweg1} B.\ D\"unweg, G.\ S.\ Grest and K.\ Kremer, to appear in the {\it conference proceedings of the IMA workshop
May 1996 at the University of Minnesota} (Springer-Verlag, Berlin)
\bibitem{duenweg2} A.\ Kopf, B.\ D\"unweg and W.\ Paul, J.\ Chem.\ Phys.\ ( in press )



\bibitem{Nose1}	S.\ Nos\'e, Prog. of Theor. Physics Supplement {\bf 103}, 1 (1991)
\bibitem{Hoover1} W.\ G.\ Hoover, Phys.\ Rev.\ A {\bf 31}, 1695 (1985)
\bibitem{Hoover2} W.\ G.\ Hoover, Phys.\ Rev.\ A {\bf 34}, 2499 (1986)
\bibitem{Tolla} D.\ Di Tolla und M.\ Ronchetti, Phys.\ Rev.\ E {\bf 48}, 1726 (1993)

\bibitem{Klein1} G.\ J.\ Martyna, Michael L.\ Klein und Mark Tuckerman, J.\ Chem.\ Phys.\
{\bf 97} (4), 2635, (1992) 









\bibitem{Siepman} J.\ I.\ Siepman and D.\ Frenkel, Molec.\ Phys.\ {\bf 75}, 59
(1992)
\bibitem{pablo1} J.\ J.\ de Pablo, M.\ Laso and U.\ W.\ Suter, J.\ Chem.\
Phys.\ {\bf 96}, 2395 (1992)


\bibitem{link} G.\ S.\ Grest, B.\ D\"unweg and K.\ Kremer, Comp.\ Phys.\ Comm.\ {\bf 55}, 269 (1989)
\bibitem{gear} C.\ W.\ Gear, {\it Initial Value Problems in Ordinary Differential Equations}
(Prentice Hall, Englewood Cliffs, 1970)

\bibitem{cicotti} S.\ Melchionna, G.\ Cicotti und B.\ L.\ Holian, Molec.\ Phys.\ {\bf 78}, 533 (1993)

\bibitem{DoiEdwards} M.\ Doi and S.\ F.\ Edwards, {\it The Theory of Polymer Dynamics} (Clarendon Press, Oxford, 1986)
\bibitem{Rouse} P.\ E.\ Rouse, J.\ Chem.\ Phys.\ {\bf 21}, 1272 (1953)
\bibitem{Li2} H.\ Z.\ Cummins, G.\ Li, W.\ M.\ Du and J.\ Hernandez, Physica A {\bf 204}, 169 (1994)
\bibitem{kaemmerer} S.\ K\"ammerer, W.\ Kob and R.\ Schilling, Phys.\ Rev.\ E (in press)

\bibitem{paul_jcp}  W.\ Paul, K.\ Binder, D.\ W.\ Heermann and K.\ Kremer, J.\ Chem.\ Phys.\ {\bf 95}, 7726 (1991)
\end{thebibliography}
\end{document}